\documentclass{revtex4}  
\usepackage[latin1]{inputenc}
\usepackage{amsmath}
\usepackage{amsfonts}
\usepackage{amssymb}
\usepackage{graphicx,color}
\usepackage[%
colorlinks=true,
urlcolor=blue,
linkcolor=blue,
citecolor=blue
]{hyperref}

\begin{document}
	
	\title{Restoring and tailoring  very high dimensional spatial entanglement of a biphoton state transmitted through a scattering medium}
	\author{Fabrice Devaux$^{1*}$, Alexis Mosset$^1$, Sébastien M. Popoff$^2$, and Eric Lantz$^1$}
	\affiliation{$^1$ Institut FEMTO-ST, D\'epartement d'Optique P. M. Duffieux, UMR 6174 CNRS \\ Universit\'e Bourgogne Franche-Comt\'e, 15b Avenue des Montboucons, 25030 Besan\c{c}on, France\\
		$^2$ Institut Langevin, ESPCI Paris, Université PSL, CNRS, 75005 Paris, France \\}
	
	\date{\today}
	\email[Corresponding author:]{fabrice.devaux@univ-fcomte.fr}
	
	
	
	\begin{abstract}
		We report experimental results where a momentum entangled biphoton state with a giant dimensionality of 8000 is retrieved and manipulated when only one photon of the pair is transmitted through a thin scattering medium. For this purpose, the transmission matrix of the complex medium is first measured with a phase-shifting interferometry measurement method using a spatial light modulator (SLM) illuminated with a laser source. From this matrix, different phase masks are calculated and addressed on the SLM to spatially control the focusing of the laser through the complex medium. These same masks are used to manipulate the phase of the biphoton wave function  transmitted by the thin diffuser in order to restore and control in the same way the momentum correlations between the far-field images of twin beams issued from strongly spatial-multi-mode spontaneous parametric down conversion.
	\end{abstract}
	\maketitle
	
	\section{Introduction}
	The control of light propagation through complex media such as multimode optical fibers, atmospheric turbulence or biological media is a major challenge for the development of protocols and systems for information processing, communication and imaging using the quantum or the classical properties of light \cite{gigan_roadmap_2021}. Besides, the manipulation of high dimensional quantum states allows a significant increase of the information that can be transmitted and processed. In this context, it is of great interest to be able to transmit high-dimensional quantum states through complex media while preserving and manipulating their dimensionality.
	
	Although it is easy to produce spatially entangled photon pairs with giant dimensionality \cite{moreau_einstein-podolsky-rosen_2014}, the entanglement is strongly degraded by the propagation in a complex medium \cite{beenakker_two-photon_2009,gnatiessoro_imaging_2019}. Several approaches have been proposed and experimentally demonstrated to counteract the effects of this environment on the transmission and propagation of quantum states. All these methods are based on the use of adaptive optics and wavefront shaping either of the pump beam \cite{lib_real-time_nodate,shekel_shaping_2021} or of the phase front of the biphoton wave function \cite{defienne_adaptive_2018}. In these works, type-I entangled photons are inseparable and therefore transmitted through the same complex medium, unlike in most practical applications. Another approach consists in using type-II entangled photons to process and detect separately the photons of a pair. In that case two methods can be considered. First, in \cite{valencia_unscrambling_2020} Valencia et al. demonstrated the transmission of  a 7-dimensional entangled state through a short commercial multi-mode fibre where unscrambling of entanglement is obtained by  manipulating only the photon that does not enter the fibre.
	
	Here we demonstrate the second method consisting in the manipulation of the photon propagating through the complex medium. Contrary to \cite{valencia_unscrambling_2020}, where the transmission matrix (TM) of the complex medium is measured using the entangled state itself, we use as in \cite{defienne_adaptive_2018} a laser source for this measurement and the design of phase masks that are applied to manipulate focusing of this laser through the complex medium. Then, we show that the same masks allow restoring and controlling  the high dimensional biphoton state transmitted through the ground glass diffuser. Instead of point detectors measuring temporal coincidences between spatially entangled photons \cite{lib_real-time_nodate,shekel_shaping_2021,valencia_unscrambling_2020}, single-photon sensitive cameras are used, allowing the detection of all photons of the images and the measurement of momentum spatial coincidences on the whole set of photons without any prior selection of the photons in time and space coincidence.
	
	\section{Experiments}
	\begin{figure}[ht]
		\centering\includegraphics[width=7cm]{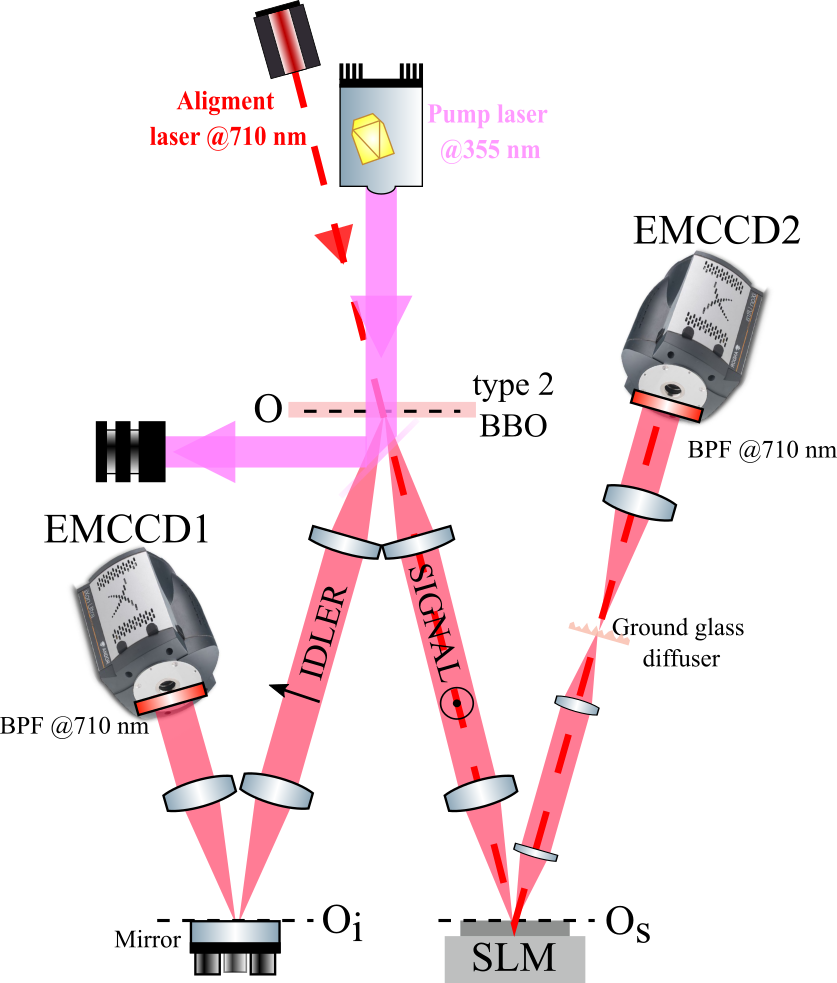}
		\caption{Experimental setup.}\label{setup}
	\end{figure}
	
	Fig. \ref{setup} shows the experimental setup. Strongly spatially multi-mode twin beams (i.e. a high-dimensional biphoton state) are generated in a non-colinear type-II geometry using  spontaneous parametric down conversion (SPDC) in a $0.8\,mm$ long $\beta$-barium borate ($\beta$-BBO) crystal pumped at $355\,nm$. The pump pulses are provided by a passively Q-switched Nd:YAG laser ($330\,ps$ FWHM pulse duration, 27 $mW$ mean power and 1 $kHz$ repetition rate). Because of the non-colinear interaction, the SPDC twin beams are separated and propagate through two different telescope systems imaging the near-field $O$ of the crystal in planes $O_s$ and $O_i$ for the signal and idler channels. For the signal, which is vertically polarized, the image plane $O_s$ is first conjugated with the SLM (Hamamatsu LCOS-SLM X10468-07). Then, the SLM plane is imaged on a ground glass diffuser  (i.e. the thin scattering medium) with a second telescope system. All telescope systems have a magnification of -1.  Finally, in order to measure momentum correlations, far-field images of the two SPDC beams are formed with $2f$ imaging systems on two separate Electron Multiplying Charge Coupled Devices (EMCCD, ANDOR iXon Ultra 897), used in photon-counting regime \cite{lantz_multi-imaging_2008}. Before detection, the photons pairs emitted around the degeneracy are selected by pass-band filters (PBF) centered at $710\,nm$ ($3\,nm$ FWHM).
	
	\begin{figure}[ht]
		\centering\includegraphics[width=13cm]{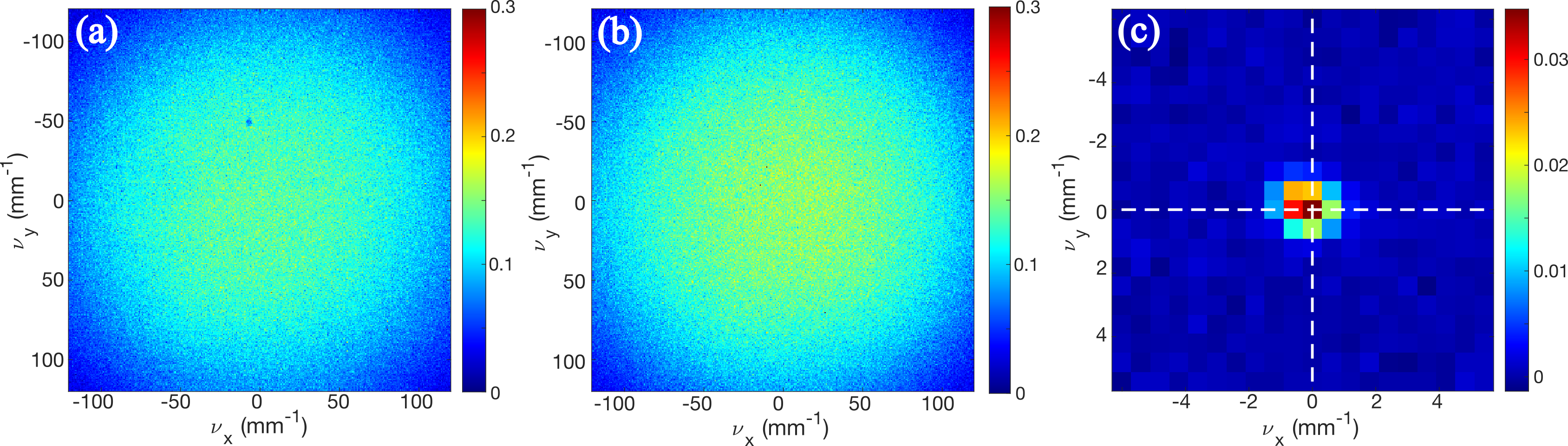}
		\caption{ Spatial distribution of the mean number of photon per pixel in far-field images of the (a) idler and (b) signal SPDC beams. (c) Normalised momentum correlation peak obtained by averaging measurements on a set of 500 twin images.}\label{correlsansMD}
	\end{figure}
	
\begin{figure}[ht]
		\centering\includegraphics[width=9cm]{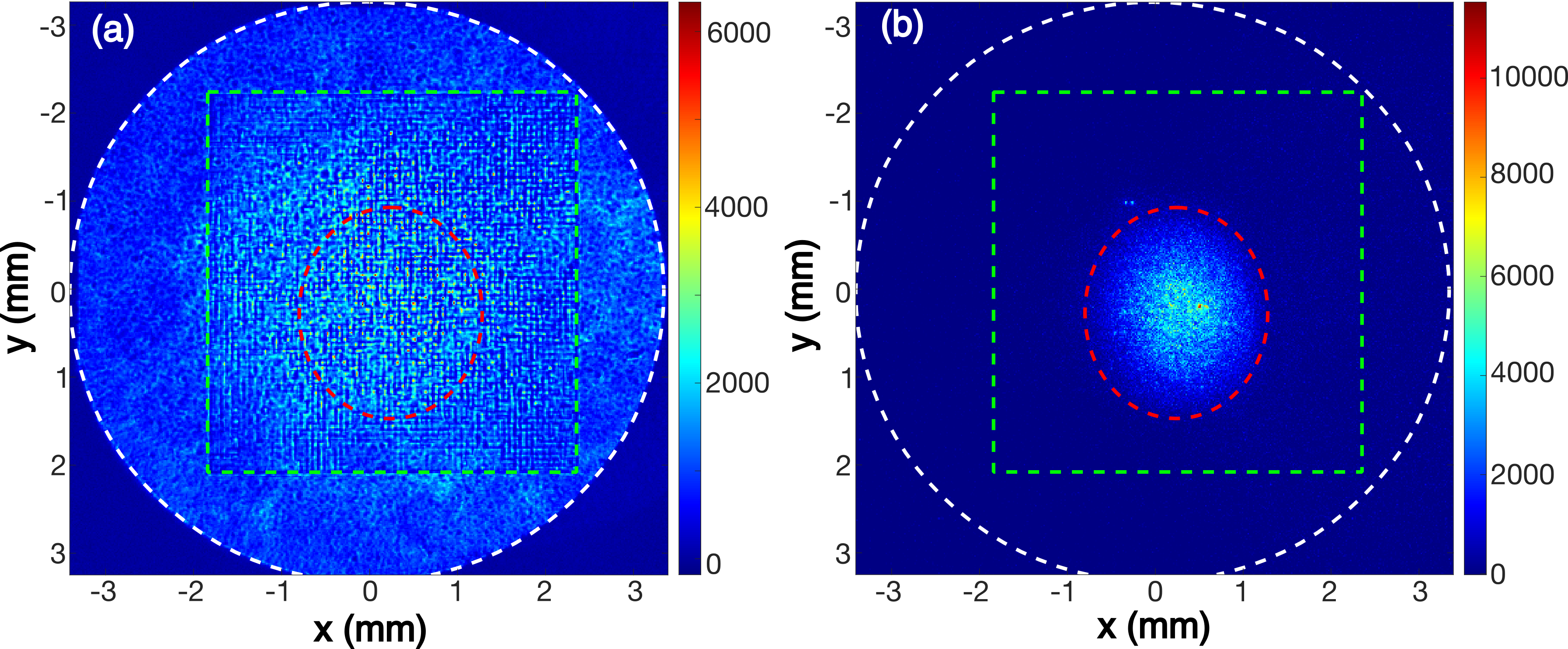}
		\caption{Near-field images graded in grayscale of the signal channel where the image planes of the crystal, the SLM and the ground glass diffuser are conjugated, (a) when the crystal is illuminated with the alignment laser, (b) when SPDC is emitted from the crystal. Dotted white circles, green squares and red ellipses denote the sizes and positions of the crystal circular mount, the phase mask addressed on the SLM  and the SPDC beam, respectively.}\label{CPHeNeSPDC}
	\end{figure}

	Fisrt, momentum correlations are measured between far-field images of twin SPDC beams without the thin diffuser and the SLM turned off. In the signal and idler images (figures \ref{correlsansMD}a and \ref{correlsansMD}b), the mean number of photon is adjusted to about 0.15 photon/pixel in order to optimize the efficiency of EMCCDs operating in photon-counting regime \cite{lantz_multi-imaging_2008}. Fig. \ref{correlsansMD}c shows the highly contrasted normalised momentum correlation peak which is obtained by averaging the measurements on a set of 500 twin images. The standard deviations of the narrow correlation peak along both axes are $\sigma_{\nu_x}=0.89\pm0.03\,mm^{-1}$ and $\sigma_{\nu_y}=0.80\pm0.03\,mm^{-1}$ and its integral, corresponding to the detection rate of photons by pairs \cite{moreau_realization_2012}, is about 23\%. From this rate, which is in a good agreement with the quantum efficiency of the whole imaging systems \cite{lantz_einstein-podolsky-rosen_2015}, we can estimate at more than 4000 the number of photons detected by pairs in a single acquisition of twin images. Moreover, the narrowness of the correlation peak demonstrates the high dimensionality of the biphoton state and the accuracy of its measurement. To complete the characterization of the biphoton state involved in our experiment, from the far-field image (Fig. \ref{correlsansMD}a) and the near-field image (Fig. \ref{CPHeNeSPDC}b) of the SPDC beams, we measure the standard deviations :  $(\sigma^{SPDC}_{\nu_x}=38.8\pm0.1\,mm^{-1},\, \sigma^{SPDC}_{\nu_y}=37.0\pm0.1\,mm^{-1})$  and $(\sigma^{SPDC}_{x}=0.707\pm0.002\,mm,\, \sigma^{SPDC}_{y}=0.796\pm0.002\,mm)$ in the momentum and position domains, respectively.  From these standard deviations, we estimate the Schmidt numbers in both dimensions \cite{law_analysis_2004,devaux_imaging_2020} : $K_x=86\pm1$ and $K_y=92\pm1$. It gives a giant spatial dimensionality of the biphoton state of $K_x\times K_y \simeq 8000$ which is good agreement with the narrowness of the measured momentum correlation peak (Fig. \ref{correlsansMD}c).
	
	A laser used for alignment (Fig. \ref*{setup}) is also used in these experiments as a coherent light source to measure the TM of the complex medium using the phase-shifting interferometry measurement method proposed in \cite{popoff_measuring_2010,popoff_controlling_2011}. Fig. \ref*{CPHeNeSPDC} shows the near-field image of the signal channel where the crystal, the SLM and the thin diffuser planes are conjugated through the telescope systems (Fig. \ref{setup}). Fig. \ref*{CPHeNeSPDC}a is obtained when the crystal is illuminated with the alignment laser and when a phase mask is addressed on the SLM. Fig. \ref*{CPHeNeSPDC}b corresponds to the near-field image of the signal SPDC beam. In both figures the white dotted circle denotes the circular mount of the BBO crystal, the green dotted square denotes the $256\times 256$ pixels phase mask addressed on the SLM and the red dotted ellipse denotes the near-field location and the size of the SPDC beam with respect to the phase mask.
	
	In the fisrt instance, the TM of the thin scattering medium is measured between a region of interest (ROI) of $256\times 256$ pixels in the SLM plane (Fig. \ref*{CPHeNeSPDC}a) and a ROI of $100\times 100$ pixels in the far-field plane imaged on the EMCCD1. Following the method proposed in \cite{popoff_controlling_2011}, we display a basis of orthogonal phase masks on the SLM and measure the output field using phase-shifting holography. To reduce the size of the TM, the pixels of the SLM are binned $4\times 4$ in order to calculate a phase mask with $64\times 64$ macropixels fitting the size of the SPDC near-field spatial modes. Once the TM has been measured, different phase masks are calculated in order to retrieve and manipulate the far-field focusing of the laser beam through the thin diffuser. To that end, we compute from the TM the masks that perform a phase conjugation operation \cite{popoff_controlling_2011}, i.e. that put in phase all the contributions from the SLM macropixels at a single or multiple localized targets in the camera plane.
	
\begin{figure}[ht]
		\centering\includegraphics[width=13cm]{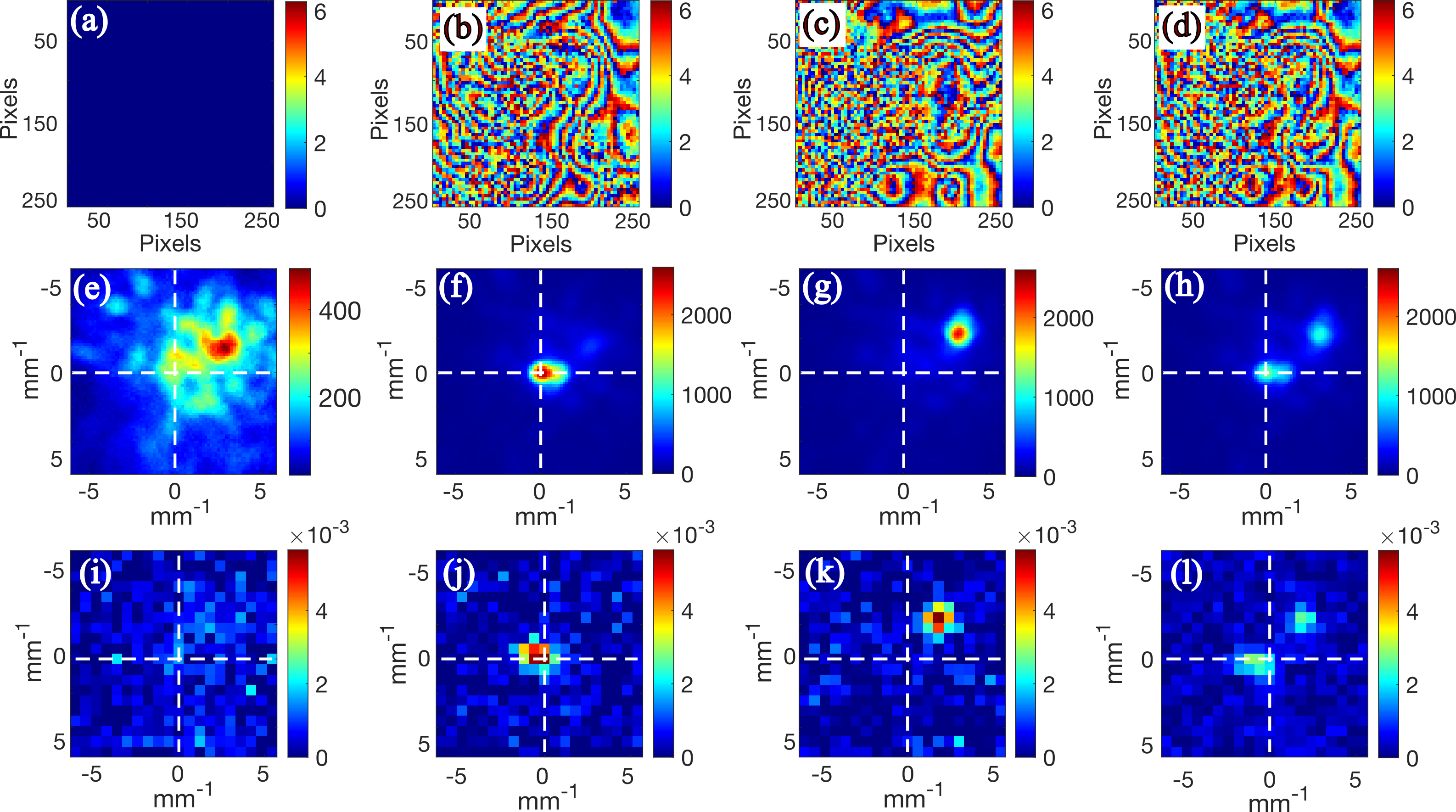}
		\caption{Experimental results. (b) to (d): $256\times 256$ pixels phase masks in radians calculated and addressed on the SLM to control the far-field focusing of the alignment laser through the ground glass diffuser (f to h) and to manipulate in the same way the momentum correlations of the biphoton state (j to l). With the SLM off (a), (e) and (i) show the one-photon and the two-photon speckle patterns observed when the alignment laser or the biphoton state are transmitted through the thin scattering medium, respectively.}\label{result}
	\end{figure}

	In Fig. \ref*{result} we show the experimental results. The first row corresponds to the different phase masks (in radians) addressed on the SLM and the second row gives the corresponding far-field image (in grayscale) of the laser intensity. The first column corresponds to the results obtained when no phase mask is addressed. In that case, the laser far-field image exhibits a broad speckle pattern due to the ground glass diffuser (Fig. \ref*{result}e). Figures \ref*{result}b and \ref*{result}c show the phase masks designed to target the far-field focusing of the alignment laser at the center and at the coordinates ($\nu_x=+3\,mm^{-1}$, $\nu_x=-3\,mm^{-1}$) of the $100\times 100$ pixels ROI on EMCCD1 (figures \ref*{result}f and \ref*{result}g). The last phase mask (Fig. \ref*{result}d) is designed to focus the laser through the scattering medium in two spots (Fig. \ref*{result}h). The third row shows the normalised momentum correlation patterns between the entangled photons obtained when the same phase masks are successively addressed on the SLM. Like in \cite{gnatiessoro_imaging_2019}, a low-contrast two-photon speckle pattern is observed (Fig. \ref*{result}i) when one photon of the biphoton state is transmitted through the  thin diffuser and when the SLM is off. When one of the phase masks is addressed, the narrow momentum correlation peak is retrieved at the same location targeted with the alignment laser (figures \ref*{result}j to \ref*{result}l). We emphasize that the phase masks used to manipulate the momentum entanglement of the biphoton state are calculated from the TM measured with the alignment laser to target its focusing. Moreover, since the TM is measured between an area on the SLM ($256\times 256$ pixels) larger than the near-field SPDC beam (Fig. \ref{CPHeNeSPDC}b), and an area on EMCCD1 ($100\times 100$ pixels)  much smaller than the far-field signal SPDC beam (Fig. \ref*{correlsansMD}b), the momentum correlations are calculated on the whole transverse section of the far-field SPDC twin beams. All momentum correlation patterns are obtained by averaging the measurements on 500 pairs of twin images. In figures \ref*{result}j and \ref*{result}k the standard deviations of the correlation peaks are about $1,0\pm 0.1\,mm^{-1}$ along the  $x$ and $y$ axes. These results show that high-dimensional momentum entanglement of the biphoton state is retrieved with a good accuracy and a good signal-to-noise ratio. However, the detection rate of photons by pairs falls down to 5\% because of additional losses induced by the transmission through the ground glass diffuser and by the low numerical aperture of the imaging system which does not allow efficient collection of the high diffraction orders produced by the small SLM pixels. With the phase mask depicted in Fig. \ref*{result}d, we obtained two narrow correlation peaks (Fig. \ref*{result}l) with amplitudes half of the correlation peak amplitude in Fig. \ref*{result}i.  
	
	\section{Conclusion}
	We have presented a method allowing the manipulation of a spatially very high dimensional biphoton state transmitted through a complex medium when only one photon of the pair is transmitted through it. The unscrambling of a 8000-dimensional entangled quantum state is evidenced by manipulating with a spatial light modulator the only photon that propagates through the medium where the used phase masks have been firstly designed to control and manipulate the focusing of a laser source through the same medium.  Although we have used a SLM and the phase-shifting interferometry method with coherent light to measure the TM of the time-stationary thin scattering medium, other methods using faster SLM devices such as a deformable mirror \cite{blochet_focusing_2017} and iterative \cite{vellekoop_phase_2008,shekel_shaping_2021}  or genetic \cite{blochet_focusing_2017}  algorithms can also be considered, opening the possibility to control in real time the spatial high dimensional entanglement of a quantum state transmitted through a more realistic dynamic complex medium such as turbulent atmosphere or biological tissues.
	
	\section*{Funding}
	This work benefited from the facilities of the SMARTLIGHT platform funded by the Agence Nationale de la Recherche (EQUIPEX+contract"ANR-21-ESRE-0040") and Région Bourgogne Franche-Comté. This work has been also supported by the EIPHI Graduate school (contract "ANR-17-EURE-0002").
	

		\bibliography{bibliomanipSLM}

\end{document}